\documentclass{ws-p8-50x6-00}

\def\cent {$\omega$\thinspace Centauri}
\def\tuca {47\thinspace Tucanae}

\def\arcmin              {$^{\prime}$} 
\def\arcm                {$^{\prime}$} 
\def\arcsec              {$^{\prime\prime}$} 
\def\arcs                {$^{\prime\prime}$} 
\def\bv                  {{\bf v}}
\def\br                  {{\bf r}}

\def\bily                {${ {10^9}}$yr}

\def\chis		 {$\chi^2$}
\def\conc                {$c$ = log ($r_t/r_c$)}
\def\deg                 {$^\circ$}

\def\ellim      {$\langle \varepsilon \rangle$}
\def\etal                {et\thinspace al.}

\def\kms                 {km\thinspace s$^{-1}$}
\def\gsim       {$\vcenter{\hbox{$>$}\offinterlineskip\hbox{$\sim$}}$}
\def\lsim       {$\vcenter{\hbox{$<$}\offinterlineskip\hbox{$\sim$}}$}

\def\lsun                {$L_{\odot}$}

\def\Mtot                {$M_{tot}$}

\def\micron              {$\mu m$}
\def\Mhr                 {$M_{hr}$}

\def\milm                {${ {10^6  M_{\odot}}}$}
\def\mily                {${ {10^6}}$yr}

\def\Mvint               {$M_V^{int}$}
\def\mlv                 {$M/L_V$}

\def\msun                {$ M_{\odot}$}

\def\mtot                {$M_{tot}$}
\def\muo                 {$\mu_{\circ}$}

\def\pmm                 {$\pm$}

\def\ra                  {$r_a$}
\def\rc                  {$r_c$}
\def\rh                  {$r_h$}
\def\rt                  {$r_t$}
\def\sig                 {$\sigma$}

\def\sigo                {$\sigma_{\circ}$}
\def\sobs                {$\sigma_{obs}$}
\def\trh                 {$t_{rh}$}

\def\voso                {$v_{\circ}/\sigma_{\circ}$}

\def\x                   {$\times$}

\begin{document}

\title{ The Internal Dynamics of Globular Clusters }

\author{ G. Meylan }

\address{ European Southern Observatory,
          Karl-Schwarzschild-Strasse 2, 
          \\D-85748 Garching bei M\"unchen, Germany 
          \\E-mail: gmeylan@eso.org}



\maketitle


\abstracts{ Galactic globular  clusters are ancient building blocks of
our Galaxy.   They represent  a  very  interesting  family of  stellar
systems in which some fundamental dynamical processes have been taking
place for more than 10 Gyr, but on time scales shorter than the age of
the  universe.  In  contrast     with galaxies, these  star   clusters
represent unique  laboratories for learning about two-body relaxation,
mass  segregation  from equipartition  of energy,  stellar collisions,
stellar  mergers, core  collapse,  and tidal disruption.  This  review
briefly summarizes some of the tremendous developments that have taken
place during the last  two decades.  It ends  with some recent results
on tidal tails around galactic globular clusters and on a very massive
globular cluster in M31. }


\section{ Introduction }

There  are about 150   globulars orbiting in  the  halo of our Galaxy.
They look like huge  swarms of  stars,  characterized by  symmetry and
apparent smoothness.  Fig.~1   below  displays an image  of   NGC~5139
$\equiv$ \cent,  the  brightest  and  most massive   galactic globular
cluster.  This 40\arcm\ by 40\arcm\ image from  the Digital Sky Survey
does not reach, in spite  of its rather  large angular size, the outer
parts of the cluster.  With its tidal  radius of about 40-50\arcm, the
apparent  diameter of \cent\ on the  plane of the sky is significantly
larger than the apparent 30\arcm\ diameter of the full moon.

Globular  clusters  are old   stellar  systems,  made of  one   single
generation of    stars.  Although   still somewhat  uncertain,   their
individual ages range   between  about 10 and 15~Gyr,   with  possible
significant  differences, up to  a few gigayears,  from one cluster to
the other.  Other properties  of globular clusters exhibit significant
variations:  e.g.,  their integrated absolute   magnitudes range  from
\Mvint\ = --1.7 to --10.1 mag; their total masses from \mtot\ = $10^3$
to 5 \x\ \milm; their galactocentric distances from 2 to 120 kpc.

\begin{figure}
\centering
\includegraphics[width=.8\textwidth]{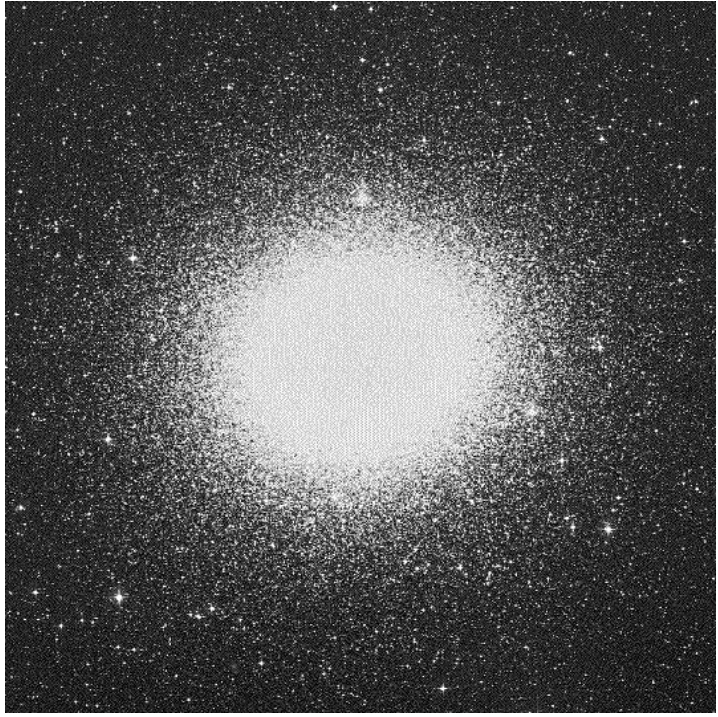}
\caption[]{NGC~5139 $\equiv$ \cent\ is the  brightest and most massive
galactic globular cluster.   This image, from  the Digital Sky Survey,
has  40\arcm\ by 40\arcm\ in  size. North is to  the  top, East to the
left.}
\label{}
\end{figure}


\section{ A few dynamical time scales }

The dynamics   of any stellar  system   may  be characterized  by  the
following three dynamical time scales: (i) the crossing time $t_{cr}$,
which is the time needed by a star to move across the system; (ii) the
relaxation time  $t_{rlx}$, which  is the time   needed by the stellar
encounters  to redistribute   energies, setting up  a  near-maxwellian
velocity distribution; (iii) the evolution time $t_{ev}$, which is the
time  during  which energy-changing mechanisms  operate, stars escape,
while the size and profile of the system change.

In the   case of globular  clusters,  $t_{cr}$ $\sim$ \mily, $t_{rlx}$
$\sim$ 100~\mily, and $t_{ev}$ $\sim$ 10~\bily. It is worth mentioning
that   several  (different  and  precise)   definitions exist  for the
relaxation time.   The most commonly used  is the half-mass relaxation
time $t_{rh}$ of Spitzer (1987,   Eq.~2-62), where the values for  the
mass-weighted mean  square velocity of the  stars and the mass density
are those evaluated at the half-mass radius of  the system (see Meylan
\& Heggie   1997 for  a review).    It  has been   suggested that  the
combination of relaxation with the chaotic nature of stellar orbits in
non-integrable potentials (e.g.,  most axisymmetric potentials) causes
a great enhancement in the rate of relaxation (Pfenniger 1986, Kandrup
\&  Willmes  1994).   Another  suggestion which,  if  confirmed, would
revolutionise  the   theory of  relaxation  was made  by  Gurzadyan \&
Savvidy (1984, 1986) who proposed a much  faster relaxation time scale
than in standard  theory, by a factor   of order $N^{2/3}$.  There  is
some support  for    this view on   observational  grounds  (Vesperini
1992a,b).
 
From size, luminosity, and mass  points of view, globular clusters are
bracketed by   open clusters on  the  lower side and  dwarf elliptical
galaxies on  the  upper side.  Table~1  displays, for   open clusters,
globular  clusters, and galaxies,  some  interesting relations between
the  above three  time  scales.   For   open clusters, crossing   time
$t_{cr}$ and  relaxation time $t_{rlx}$ are  more  or less equivalent,
both being significantly   smaller than the evolution  time  $t_{ev}$.
This means  that most open  clusters dissolve within a  few gigayears.
For  galaxies, relaxation time  $t_{rlx}$  and evolution time $t_{ev}$
are more or less equivalent, both  being significantly larger than the
crossing   time $t_{cr}$.  This means  that  galaxies are not relaxed,
i.e., not dynamically evolved.  It  is only for globular clusters that
all three time scales  are significantly different, implying plenty of
time for a significant  dynamical evolution in these  stellar systems,
although avoiding quick evaporation.

Consequently,   globular clusters  represent   an interesting class of
dynamical stellar systems in which some dynamical processes take place
on  time scales shorter than their  age, i.e., shorter than the Hubble
time, providing us  with unique  dynamical laboratories  for  learning
about  two-body relaxation,  mass  segregation  from  equipartition of
energy,  stellar collisions, stellar mergers,  and core collapse.  All
these   dynamical phenomena are   related to   the internal  dynamical
evolution only,  and would also  happen in isolated globular clusters.
The external dynamical   disturbances  ---  tidal  stripping   by  the
galactic gravitational  field    --- influence  equally   strongly the
dynamical evolution of globular clusters.

\begin{table}[t]
\caption{}
\begin{center}
\footnotesize
\begin{tabular}{|l|c|c|c|l|}
\hline  
 & & \\
open clusters &t$_{cr}$ $\sim$ t$_{rlx}$ $\ll$ t$_{ev}$ & quickly dissolved\\
 & & \\
globular clusters & t$_{cr}$ $\ll$  t$_{rlx}$ $\ll$  t$_{ev}$ & \\
 & & \\
galaxies          & t$_{cr}$ $\ll$  t$_{rlx}$ $\sim$ t$_{ev}$ &not relaxed\\
 & & \\
\hline
\end{tabular}
\end{center}
\end{table}


\section{ Model building for globular clusters }

Already before the pioneering work of von Hoerner (1960), who made the
first  $N$-body calculations  with $N$   =  16, it was realized   that
computation  of   individual stellar motions    could be  replaced  by
statistical  methods.  Some parallels   were drawn between a molecular
gas  and star  clusters:  the  stars  were considered  as  mass points
representing  the   molecules in  a collisionless   gas.   The analogy
between  a gas  of   molecules  and a  gas   of  stars is  subject  to
criticisms, since the mean free path of  a molecule is generally quite
small  compared with the size  or scale height  of the system, whereas
the mean free path of  a star is much larger  then the diameter of the
cluster;  in addition  molecules  travel  along straight lines,  while
stars  move along orbits  in  the gravitational  potential of all  the
other  stars of  the  stellar system.  Stellar  collisions in clusters
were studied   by  Jeans (1913), who   remarked  that  they   might be
important in such stellar  systems. The problem  was then to seek  the
possible spherical distribution of such a gas in a steady state.

\subsection{ Boltzmann's equation }

The commonest way of defining a model of a star cluster is in terms of
its distribution function   $f(\br,\bv,m)$, which  is defined by   the
statement that $fd^3\br  d^3\bv dm$ is the  mean number  of stars with
positions in a small box $d^3\br$ in space,  velocities in a small box
$d^3\bv$ and masses in an interval $dm$.  In terms of this description
a fairly  general equation for  the dynamical evolution is Boltzmann's
equation,
$$
{\partial f\over\partial t} + \bv.\nabla_{\br}f -
\nabla_{\br}\Phi.\nabla_{\bv}f = {\partial f\over\partial t}_{enc}, \eqno(1)
$$ 
where $\Phi$ is the smoothed   gravitational potential per unit  mass,
and  the right-hand side describes  the effect of two-body encounters.
The distribution  $f$ is a  function of 7  variables  if we  take into
account time.   This is rather  more than can  be  handle.  But  it is
possible to reduce  the  complexity posed by Boltzmann's  equation  by
taking moments.

By  taking moments    of the Boltzmann's  equation    with  respect to
velocities we obtain,  for n = 0 and  1, the Jeans equations which are
expressions describing the rotation and the velocity dispersion:
$$
\int Boltzmann \cdot v_j^n  d^3{\bf v}  = Jeans Equ.  \eqno(2)
$$

By taking moments of the Jeans equations with respect to positions, we
obtain the Tensor Virial equations  which are expressions relating the
global kinematics to  the morphology of  the  system, e.g., the  ratio
\voso\ of ordered to random motions:
$$
\int Jeans \cdot x_j^n  d^3{\bf x}  = Tensor~ Virial  \eqno(3)
$$

In these ways,  we obtain information about  the general properties of
solutions of Boltzmann's equation without recovering any solutions.

\subsection{ Liouville's equation and Jean's theorem }

The general  Boltzmann's equation can be  greatly  simplified in other
ways.  Because $t_{cr}$ is so short, after a  few orbits the stars are
mixed into a nearly stationary distribution, and so the term $\partial
f/\partial t$ is practically equal to zero.  In a similar way, because
\trh\  is so long, the collision  term $(\partial f/\partial t)_{enc}$
can be ignored.  What is left, i.e.,
$$
\bv.\nabla_{\br}f - \nabla_{\br}\Phi.\nabla_{\bv}f = 0, \eqno(4)
$$
is   an equilibrium form of  what    is frequently called  Liouville's
equation,  or the  collisionless  Boltzmann's equation, or the  Vlasov
equation.

In simple  cases, the general  solution  of Equ.~4  is given by  Jeans'
theorem, which states that $f$ must be  a function of the constants of
the equations of  motion of a star,   e.g., of the  stellar energy per
unit  mass  $\varepsilon =  v^2/2 + \Phi$.    Such quantities are also
called integrals of  the motion.  If not  all integrals of the  motion
are known,  such functions  are  still solutions, though  not the most
general.  For a  self-consistent  solution, the distribution  function
$f$ must correspond to the   density  $\rho$ required to provide   the
cluster potential $\Phi_c$, i.e.:
$$
\nabla^2\Phi_c = 4\pi G\rho 
               = 4\pi G\int mfd^3\br d^3\bv dm.  
    \eqno(5)
$$
Many different kinds of models  may be constructed with this approach.
In the first place there is considerable  freedom of choice over which
integrals of the motion to  include.  In the  second place one is free
to   choose the functional  dependence of  these  integrals, i.e., the
analytic  form of the distribution function   (see, e.g., Binney 1982,
and Binney \& Tremaine 1987).

King (1966)  provided   the  first  grid of   models  (with  different
concentrations  \conc\ where  \rt\ and   \rc\ are the   tidal and core
radii,  respectively)  that    incorporate the three   most  important
elements  governing globular cluster structure: dynamical equilibrium,
two-body relaxation, and tidal truncation.  These models depend on one
integral  of  the motion only  ---  the stellar  energy  per unit mass
$\varepsilon$ ---  and  the  functional  dependence is based    on the
lowered maxwellian  (see Equ.~6 below).  Such  models are spherical and
their velocity dispersion tensor is everywhere isotropic.

Models  more complicated  have been   built  since then.  Da Costa  \&
Freeman  (1976) generalized the   simple  single-mass King  models  to
produce more realistic  multi-mass  models with full equipartition  of
energy in the  centre.   Gunn \& Griffin (1979)  developed  multi-mass
models  whose distribution functions  depend on the stellar energy per
unit mass  $\varepsilon$ and the specific  angular momentum $l$.  Such
models are spherical and have a radial anisotropic velocity dispersion
($\overline {v_r^2}$ $\not=$  $\overline {v_{\theta}^2}$  = $\overline
{v_{\phi}^2}$).  Called King-Michie models, they associate the lowered
maxwellian of  the   King model  with  the  anisotropy factor  of  the
Eddington models:
$$ f(\varepsilon,l) \propto 
(\exp(-2j^2\varepsilon)-\exp(-2j^2\varepsilon_t)) ~
\exp(-j^2 l^2/r_a^2) \eqno(6) 
$$
Lupton \& Gunn  (1987) developed multi-mass models  whose distribution
functions depend on a third  integral of motion  $I_3$, in addition to
the  stellar energy per unit mass  $\varepsilon$ and  the component of
angular  momentum parallel to   the rotation axis $l_z$.   Although no
general   analytical form  for  a  third  integral  is  available, the
existence of an  analytic third  integral  of motion $I_3$ in  special
cases has been known   for decades, since   the work by  Jeans (1915).
Because the rotation creates a non-spherical  potential, $I_3$ = $l^2$
is in   fact only  an  approximate  integral  and  Lupton  \&   Gunn's
distribution  function  does not  obey  the  collisionless Boltzmann's
equation for equilibrium (Eq.~4).

\begin{table}[t]
\caption{Dynamical models of globular star clusters}
\begin{center}
\footnotesize
\begin{tabular}{|l|l|c|c|c|c|c|c|}
\hline
&& \multicolumn{3}{|c|}{~}& \multicolumn{3}{|c|}{~} \\
&& \multicolumn{3}{|c|}{Static Models}& \multicolumn{3}{|c|}{Evolutionary Models} \\
&& \multicolumn{3}{|c|}{~}& \multicolumn{3}{|c|}{~} \\
\cline{3-8}
& &~~ & & & & & \\
&          & King & Michie- & 3-Integral & Gas & Fokker- & N-Body \\
&          &      &  King   &            &     & Planck  &        \\
& &~~ & & & & & \\
\hline
\multicolumn{8}{|l|}{Dynamical Features }  \\
\hline
& &~~ & & & & & \\
 Anisotropy& & ... & $\surd$ &  $\surd$ & $\surd$ & $\surd$ & $\surd$ \\
 Rotation  & & ... &  ...    &  $\surd$ &  ...    & $\surd$ & $\surd$ \\
 Flattening& & ... &  ...    &  $\surd$ &  ...    & $\surd$ & $\surd$ \\
& &~~ & & & & & \\
\hline
\multicolumn{8}{|l|}{Dynamical Processes }  \\
\hline
& &~~ & & & & & \\
Stellar  &1-body& ...   & ...   & ...   &$\surd$&$\surd$&$\surd$\\
evolution&      &       &       &       &       &       &       \\
& &~~ & & & & & \\
Relaxation&2-body&$\surd$&$\surd$&$\surd$&$\surd$&$\surd$&$\surd$\\
& &~~ & & & & & \\
Tidal     &      &       &       &       &       &       &       \\
Interactions&2-body&... & ...   & ...   & ...   &$\surd$&$\surd$\\
Collisions&      &       &       &       &       &       &       \\
& &~~ & & & & & \\
 Stellar&2-body&$\surd$& ...   & ...   & ...   &$\surd$&$\surd$\\
Escape  &      &       &       &       &       &       &       \\
& &~~ & & & & & \\
Primordial&3- and& ...   & ...   & ...   &$\surd$&$\surd$&$\surd$\\
Binaries  &4-body&       &       &       &       &       &       \\
& &~~ & & & & & \\
Stellar&collision-&$\surd$&$\surd$&$\surd$&$\surd$&$\surd$&$\surd$\\
Motions&less      &       &       &       &       &       &       \\
& &~~ & & & & & \\
Steady&collision-&$\surd$&$\surd$&$\surd$& ...   &$\surd$&$\surd$\\
Tide&less        &       &       &       &       &       &       \\
& &~~ & & & & & \\
Disk&collision-& ...   & ...   &  ...  & ...   &$\surd$&$\surd$\\
Shocking&less  &       &       &       &       &       &       \\
& &~~ & & & & & \\
\hline
\end{tabular}
\end{center}
\end{table}

These were notable landmarks in these developments, among many others.
Table~2  hereafter, from Meylan \&  Heggie (1997), list for the static
models (King, King-Michie, 3-Integral) and for the evolutionary models
(gas, Fokker-Planck,   N-Body) the dynamical  features  and  dynamical
processes   they  take into  account.   Under   the heading  Dynamical
Process, the  second column in Table~2   states what kind  of physical
process it is that is named in the first column.


\section{Parametric and non-parametric approaches}

The method in the above section for analyzing globular cluster data is
a model-building, or  parametric, approach.  One begins by postulating
a  functional   form for  the   distribution   function  $f$   and the
gravitational potential $\Phi$; often the two are linked via Poisson's
equation, i.e.  the stars described by $f$  are assumed to contain all
of the mass  that contributes to $\Phi$.   This $f$ is then  projected
into observable space and its predictions compared with  the data.  If
the discrepancies are  significant, the model  is rejected and another
one is tried.   If no combination  of functions $\{f,\Phi\}$ from  the
adopted  family can be found  that reproduces the  data, one typically
adds  extra  degrees of freedom until   the  fit is satisfactory.  For
instance, $f$ may be allowed to depend on a larger number of integrals
of the   motion (Lupton  \&   Gunn 1987) or    the range  of  possible
potentials may be  increased by postulating  additional populations of
unseen stars (Da Costa \& Freeman 1976).

This approach has enjoyed  considerable popularity, in part because it
is computationally  straightforward but also  because, as  King (1981)
has emphasized, globular cluster  data  are generally well fitted   by
these  standard models.  But one never  knows which of the assumptions
underlying the models are adhered to by the  real system and which are
not.  For instance, a deviation between the surface density profile of
a globular cluster and the profile predicted by  an isotropic model is
sometimes taken as evidence that the real cluster is anisotropic.  But
it  is equally possible that  the adopted form for $f(\varepsilon)$ is
simply  in error,  since by   adjusting   the dependence  of $f$    on
$\varepsilon$ one   can    reproduce   any  density   profile  without
anisotropy.  Even  including  the additional constraint of  a measured
velocity dispersion profile does not  greatly improve matters since it
is  always  possible  to  trade off  the   mass distribution with  the
velocity anisotropy in such a way as to leave the observed dispersions
unchanged (Dejonghe  \&   Merritt 1992).  Conclusions  drawn  from the
model-building studies are hence very difficult to interpret; they are
valid only to the extent that the assumed functional forms for $f$ and
$\Phi$ are correct.

These arguments  suggest  that  it  might  be profitable to  interpret
kinematical data from  globular   clusters in an  entirely   different
manner, placing much stronger demands on  the data and making fewer ad
hoc assumptions about $f$ and  $\Phi$.  Ideally, the unknown functions
should  be    generated non-parametrically from   the  data.   Such an
approach pioneered  by Merritt (see,  e.g., Merritt 1993a,b, 1996) has
rarely been  tried in the past  because of the inherent instability of
the deprojection  process.  We provide here  after the  results of two
studies (parametric  and non-parametric, respectively) of the globular
cluster \cent, both studies using  exactly the same observational data
(surface brightness profile and stellar radial velocities).

\subsection{Parametric approach applied to \cent}

The mean  radial velocities obtained with  CORAVEL (Mayor \etal\ 1997)
for  469 individual stars  located   in the galactic  globular cluster
\cent\  provide  the  velocity   dispersion  profile.   It   increases
significantly from the outer  parts  inwards: the 16 outermost  stars,
located   between 19.2\arcm\ and 22.4\arcm\  from   the center, have a
velocity dispersion \sig\ = 5.1 \pmm\ 1.6 \kms, while the 16 innermost
stars,   located  within 1\arcm\  from the   center,  have  a velocity
dispersion  \sig\ = 21.9  \pmm\ 3.9 \kms.   This  inner value of about
\sigo\ = 22 \kms\ is the largest velocity dispersion value obtained in
the core of any galactic globular cluster (Meylan \etal\ 1995).

A  simultaneous fit  of these   radial  velocities and of the  surface
brightness  profile to    a  multi-mass King-Michie  dynamical   model
provides mean estimates of the total mass equal to \Mtot\ = 5.1 \milm,
with  a corresponding  mean  mass-to-light  ratio \mlv\  =  4.1.   The
present results   emphasize  the fact  that  \cent\  is  not  only the
brightest but also, by far, the most massive galactic globular cluster
(Meylan \etal\ 1995).

The  fact  that only models with   strong anisotropy  of  the velocity
dispersion (\ra\ = 2-3 \rc) agree with  the observations does not give
a definitive proof of   the presence of   such anisotropy  because  of
fundamental  indetermination  in  the  comparison between  King-Michie
models and observations.  A strong anisotropy is nevertheless expected
outside of the core of  \cent, given the large  value of the half-mass
relaxation  time  of about  26 $\leq$  \trh  $\leq$ 46  \bily) (Meylan
\etal\ 1995).

The reliability of the present application of King-Michie models might
be  questionable  on a few fundamental   points.  In  addition  to the
arbitrary   choices  of the two  integrals  of  the  motion and of the
functional dependence    of  the distribution function  on   these two
integrals, there is  also the assumption  of thermal equilibrium among
the different mass classes in the central  parts of the cluster.  From
a theoretical point  of  view, mass  segregation has been  one  of the
early important   results to emanate  from   small N-body simulations.
Since    then, large N-body  simulations   and  models integrating the
Fokker-Planck equation    for   many thousands of   stars   have fully
confirmed  the presence of equipartition.   Thanks to the high angular
resolution  of  the Hubble  Space   Telescope (HST) cameras  (FOC  and
WFPC2), mass segregation has  now been observed  in the core of  a few
galactic globular clusters (see,  e.g., Anderson 1997, 1999).  In  the
case  of \tuca, the observed  luminosity  function by Anderson  (1997,
1999)  is  in close agreement  with equipartition-assuming King-Michie
models and fails to fit the  no-segregation models.  This dichotomy is
not as clear in the case of \cent, probably because of its rather long
central relaxation time.

The problem about mass segregation does not  concern its existence ---
it is happening ---, but  rather its quantitative evolution. Can there
be an  end  to mass segregation, i.e.,   does the system ever  reach a
stable  thermal equilibrium~?    Underlying  is the  problem  of  core
collapse (see  Spitzer  1969, Chernoff  \&   Weinberg 1990),  which is
briefly described in \S~7 below.

\subsection{Non-parametric approach applied to \cent}

The  stellar dynamics  of  \cent\ is  inferred   from the same  radial
velocities of  469 stars used   in  \S~4.1 (Mayor  \etal\  1997).   By
assuming that the residual velocities are  isotropic in the meridional
plane, $\sigma_{\varpi}=\sigma_z\equiv\sigma$, Merritt \etal\   (1997)
derived the dependence  of  the two independent  velocity  dispersions
$\sigma$ and  $\sigma_{\phi}$ on various  positions in  the meridional
plane.  The  central  velocity dispersion parallel   to the meridional
plane is \sigo\ = $17^{+2.1}_{-2.6}$ \kms.   With this approach, there
is  no evidence for significant  anisotropy  anywhere in \cent.  Thus,
this    cluster can reasonably be    described  as an isotropic oblate
rotator (Merritt \etal\ 1997).

\begin{figure}
\centering
\includegraphics[width=.7\textwidth, angle=-90]{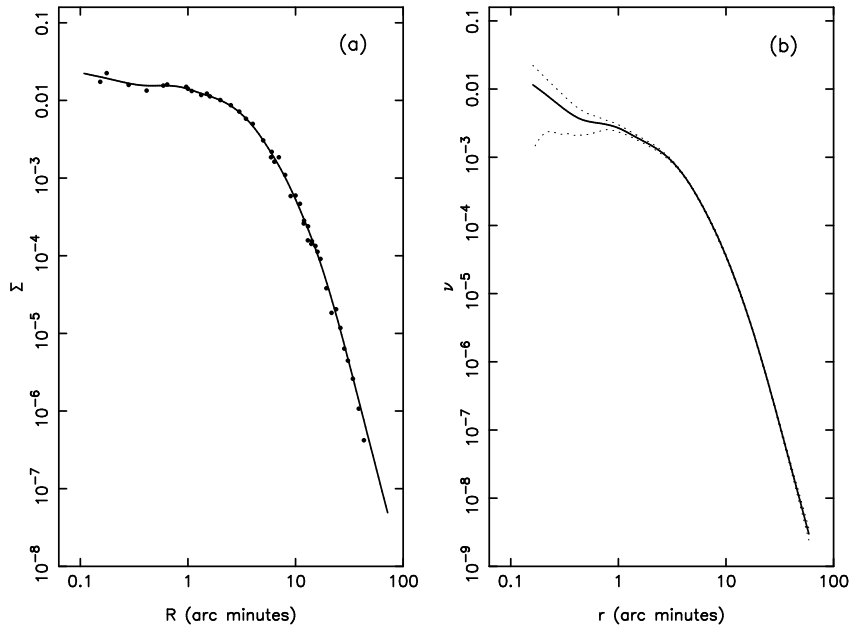}
\caption[]{Surface brightness  (a) and space  density (b) profiles for
\cent\ from Merritt \etal\ (1997). }
\label{}
\end{figure}

The  binned  surface brightness  measurements  from Meylan  (1986) are
plotted in Fig.~2a,  where  the solid line in   is an estimate of  the
surface  brightness profile  $\Sigma(R)$,  as   the solution  to   the
optimization problem.  The  estimate of space density profile $\nu(r)$
may be defined as the Abel inversion of the estimate $\Sigma(R)$:
$$
\nu(r) = -{1\over\pi}\int_r^{\infty} {d\Sigma\over dR} 
{dR\over\sqrt{R^2-r^2}}.  \eqno(7)
$$
The dashed lines in Fig.~2b are 95\%  confidence bands on the estimate
of $\nu(r)$.  Here $r$  is an azimuthally-averaged  mean radius.  Both
profiles  are normalized to unit total  number.  This profile actually
has a power-law cusp, $\nu\sim r^{-1}$,  inside of 0.5\arcmin; however
the  confidence bands are consistent  with a wide   range of slopes in
this region, including even a profile that declines toward the center.

The  gravitational   potential and mass   distribution   in \cent\ are
consistent  with  the predictions  of a  model   in which the mass  is
distributed in the same way as the bright  stars, although the cluster
is assumed to be oblate and edge-on but  mass is not assumed to follow
light.  The  central  mass density is $2110^{+530}_{-510}M_{\odot}{\rm
pc}^{-3}$.   However   this result may   be  strongly dependent on the
assumption that the velocity  ellipsoid is isotropic in the meridional
plane.  This central mass  density determination is in full  agreement
with   the values deduced  from King-Michie  models   by Meylan \etal\
(1995).

There is no significant evidence for a difference between the velocity
dispersions parallel and  perpendicular to the  meridional plane.  The
mass  distribution inferred  from    the kinematics is  slightly  more
extended  than, though not  strongly inconsistent with, the luminosity
distribution.    The   derived   two-integral   distribution  function
$f(\varepsilon,l_z)$ for the stars  in \cent\ is fully consistent with
the available data.

Large amount of kinematical data (radial velocities and proper motions
for a  few thousand stars) will  soon allow the  efficient  use of the
non-parametric   approach  in the  case  of the  largest  two galactic
globular clusters,  viz.  \cent\ and   \tuca\ (Freeman \etal,   Meylan
\etal, both in preparation).


\section{ Systemic rotation of \cent }

Systemic rotation  in globular clusters has  been  expected for a long
time, especially in \cent, because of its significant flattening.  The
first clear  evidence of such rotation was   observed, in this cluster
and in \tuca, by Meylan  \& Mayor (1986).   More recently, rather than
fitting the data to a  family of models, estimate  of the rotation was
obtain non-parametrically, by direct operation on  the data by Merritt
\etal\ (1997).

\begin{figure}
\centering
\includegraphics[width=.7\textwidth]{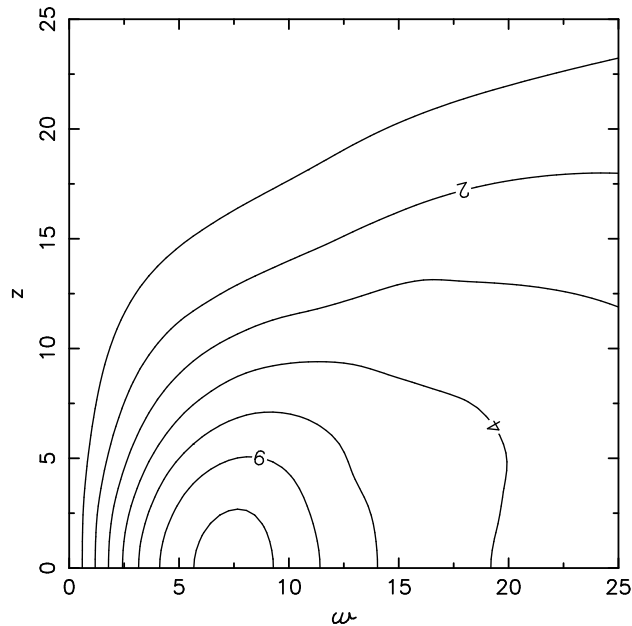}
\caption[]{Rotation in NGC~5139  $\equiv$ \cent: estimate  of the mean
azimuthal velocity $\overline{v}_{\phi}$   in the meridional plane  of
\cent, displayed  here in the  North-West quadrant.   Distances are in
arc  minutes and contours are  labeled in  \kms.  From Merritt \etal\
(1997). }
\label{}
\end{figure}

Fig.~3 displays the  contours of constant  $\overline{v}_{\phi}$ which
are  remarkably similar in  shape  to  those of  the parametric  model
postulated by Meylan \& Mayor (1986), at least  in the region near the
center where the  solution is strongly constrained  by the  data.  The
rotational velocity   field  is  clearly  not  cylindrical;   instead,
$\overline{v}_{\phi}$   has a peak value   of 8~\kms\ at about 7\arcm\
from  the  center in the   equatorial plane, and  falls off  both with
increasing  $\varpi$  and $z$.  In the   region  inside the  peak, the
rotation is  approximately solid-body;  at large radii,  the available
data do  not  strongly constrain the form  of  the rotational velocity
field.  The  mean  motions are consistent  with axisymmetry  ,  once a
correction is made for perspective rotation resulting from the cluster
proper motion.
 
The  above inferred  rotational    velocity  field in   \cent\   agree
remarkably with the predictions of the theoretical  model by Einsel \&
Spurzem (1999), who have investigated the influence of rotation on the
dynamical evolution of  collisional  stellar  systems by  solving  the
orbit-averaged              Fokker-Planck         equation          in
$(\varepsilon,l_z)$-space. However it is  not clear that  any relevant
comparison exists:  because of the long  relaxation time in \cent, the
rotation probably still  reflects to a  large extent the  state of the
cluster shortly  after   its  formation.  The   observed  estimate  of
$\overline{v}_{\phi}(\varpi,z)$  might therefore be  most  useful as a
constraint on cluster formation models.

But  things may  be  even  more   complicated~!  Using their   calcium
abundances for about 400 stars with  radial velocities by Mayor \etal\
(1997), Norris \etal\ (1997)  found that the  20\% metal-rich  tail of
the  [Ca/H] distribution is not only  more centrally concentrated, but
is also  kinematically   cooler than  the 80\%  metal-poor  component.
While the   metal-poorer  component  exhibits  well-defined   systemic
rotation,   the  metal-richer   one   shows no   evidence    of it, in
contradistinction  to the simple  dissipative  enrichment scenario  of
cluster formation.


\section{ Rotation vs. velocity dispersion }

All results  about  rotation depend on   the  value of the   angle $i$
between the plane of the sky and the axis  of symmetry of the cluster.
This  angle remains unknown.  Since   the  two best studied  clusters,
viz. \cent\  and \tuca, belong  to the small  group of clusters which,
among the 150  galactic globular clusters,  are the flattest ones,  we
can expect,  from a statistical point of  view,  that their angles $i$
should not be very different from 0\deg\ $\leq$ $i$ $\leq$ 30\deg, the
clusters   being seen  nearly  edge-on.    The importance of  rotation
(namely,  of its projection along the  line of sight) increases as $i$
gets closer to 0\deg.

The relative importance of rotational   to random motions is given  by
the ratio \voso, where $v_{\circ}^2$  is the mass-weighted mean square
rotation  velocity  and  $\sigma_{\circ}^2$  is the mass-weighted mean
square random  velocity.  For $i$ =  90\deg\ and 60\deg, in \cent\ the
ratio \voso\ = 0.35 and 0.39 and in \tuca\ the ratio \voso\ = 0.40 and
0.46, respectively (Meylan  \& Mayor 1986).   Even with  $i$ = 45\deg,
the  dynamical importance of rotation  remains weak compared to random
motions.  The   ratio of  rotational  to  random  kinetic  energies is
$\simeq$ 0.1, confirming the  fact that  globular clusters are,  above
all, hot stellar systems.

Rotation  has been directly observed  and  measured in twelve globular
clusters  (see  Table~7.2  in Meylan \&   Heggie   1997).  The diagram
(\voso~vs.~\ellim), of the  ratio  of ordered  $v_{\circ}$ to  random
$\sigma_{\circ}$ motions as a function of  the ellipticity \ellim, has
been frequently used   for  elliptical galaxies   and  its meaning  is
extensively  discussed in Binney  \& Tremaine (1987 Chapter~4.3).  The
low  luminosity ($L$ \lsim\  2.5 10$^{10}$  \lsun) elliptical galaxies
and spheroids have (\voso,\ellim) values which are scattered along the
relation for   oblate   systems  with   isotropic  velocity-dispersion
tensors,  while the high luminosity ($L$   \gsim\ 2.5 10$^{10}$ \lsun)
elliptical  galaxies have  (\voso,\ellim) values   which are scattered
below the  above  relation,  indicating the  presence  of  anisotropic
velocity-dispersion  tensors.   Given their  small  mean ellipticities
(0.00 $\leq$ \ellim\  $\leq$ 0.12), globular  clusters are located  in
the  lower-left corner of  the  (\voso\ vs.   \ellim) diagram, an area
characterized   by     isotropy     or mild    anisotropy      of  the
velocity-dispersion tensor.


\section{ Overwhole dynamical evolution towards core collapse }

Till the  late ninety seventies, globular clusters  were thought to be
relatively   static   stellar systems  since  most  surface-brightness
profiles of globular  clusters were successfully fitted by equilibrium
models.  Nevertheless,   it had been already    known, since the early
sixties, that globular clusters  had to evolve dynamically,  even when
considering    only  relaxation,   which  causes   stars  to   escape,
consequently cluster  cores to contract and envelopes  to expand.  But
dynamical evolution  of globular   clusters  was not yet  a   field of
research by itself, since the very  few theoretical investigations had
led  to   a most   puzzling  paradox:  core  collapse  (H\'enon  1961,
Lynden-Bell \& Wood 1968).

It was  only in the early eighties  that the  field grew dramatically.
On  the   theoretical side,  the development   of high-speed computers
allowed numerical  simulations   of  dynamical  evolution.   Nowadays,
Fokker-Planck and  conducting-gas-sphere evolutionary models have been
computed well into core collapse  and beyond, leading to the discovery
of  possible post-collapse oscillations.  In  a  similar way, hardware
and software  improvements  of N-body  codes provide  very interesting
first results for 10$^4$-body  simulations (Makino 1996a,b, Spurzem \&
Aarseth 1996, Portegies Zwart \etal\ 1999), and give the first genuine
hope, in   a  few years,    for   10$^5$-body simulations.   On    the
observational side, the   manufacture   of   low-readout-noise  Charge
Coupled Devices (CCDs), combined  since   1990 with the high   spatial
resolution of   the    Hubble Space   Telescope   (HST), allow    long
integrations on faint    astronomical targets in  crowded fields,  and
provide improved data analyzed with sophisticated software packages.

\section{ Gravothermal instability, gravothermal oscillations }

For  many  years (between about 1940   and 1960)  secular evolution of
globular cluster  was understood in terms  of the evaporative model of
Ambartsumian (1938)  and Spitzer (1940).   In this model it is assumed
that two-body relaxation attempts to  set up a maxwellian distribution
of velocities on the time  scale of a relaxation  time, but that stars
with velocities above the escape  velocity promptly escape.  The  next
major step in understanding came when it was discovered that evolution
arises also when stars escape  from the inner parts  of the cluster to
larger radii, without necessarily escaping altogether.  Antonov (1962)
realised   that these  internal   readjustments need  not  lead  to  a
structure in  thermal equilibrium, because  thermal equilibrium may be
unstable in  self-gravitating systems (see  Lynden-Bell \& Wood 1968).
The  well   known  process of  core   collapse  is  interpreted   as a
manifestation of the gravothermal instability.

Core collapse has been first observed and studied in simulations using
gas  and  Fokker-Planck models.  For  an isolated  cluster  (without a
tidal field) the time scale for the entire evolution of the core (when
the density  has formally become  infinite) is about 15.7 $t_{rh}$(0),
when expressed in terms of the initial half-mass relaxation time (Cohn
1980).  This result is  for an isotropic code  starting from a Plummer
model with stars of equal mass, while for an anisotropic code the time
extends to 17.6 $t_{rh}(0)$ (Takahashi 1995).

The  collapse time is generally  shorter  in the  presence of  unequal
masses (Inagaki  \& Wiyanto 1984,  Chernoff \& Weinberg 1990).  Murphy
\& Cohn (1988)  give   surface  brightness and  velocity    dispersion
profiles at various   times   during collapse, for  a   system  with a
reasonably realistic present-day  mass spectrum.   Addition of effects
of stellar evolution, modeled as instantaneous mass loss at the end of
main sequence evolution, delays the onset  of core collapse (Angeletti
\& Giannone  1980,  Applegate 1986,   Chernoff \&  Weinberg 1990,  Kim
\etal\ 1992).  The effect of a galactic time-dependent tidal field can
be to accelerate core collapse (Spitzer \& Chevalier 1973).

Examples of $N$-body  models which illustrate  various aspects of core
collapse include Aarseth (1988), where  $N = 1,000$, Giersz \&  Heggie
(1993) ($N\le2,000$), Spurzem \&  Aarseth  (1996) ($N =  10,000$), and
Makino (1996a,b; see Fig.~4 hereafter) ($N\le32,000$).

\begin{figure}
\centering
\includegraphics[width=1.0\textwidth]{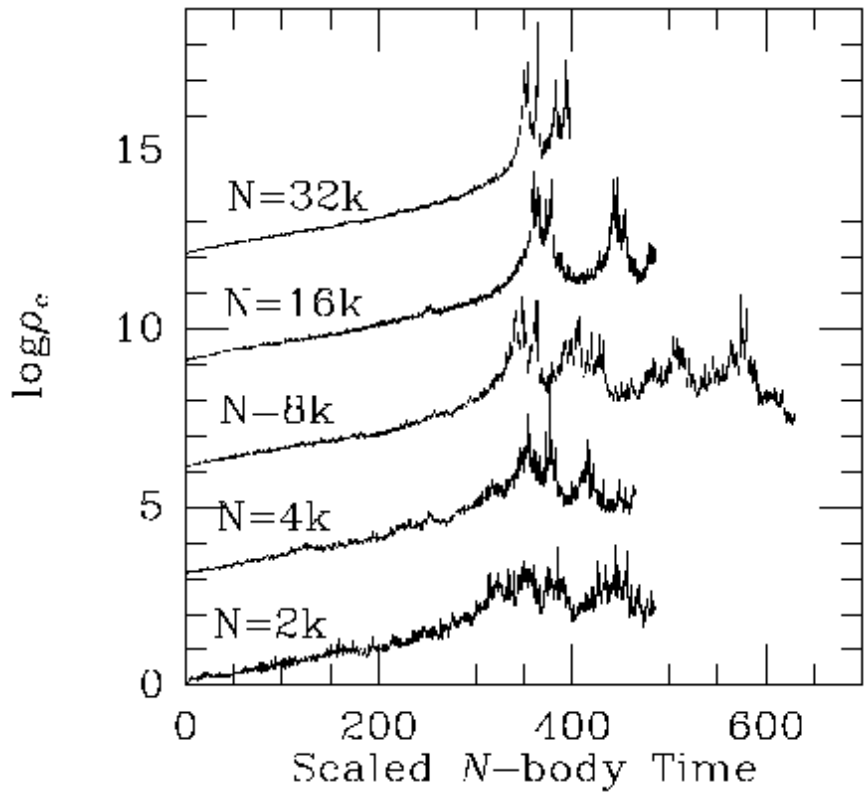}
\caption[]{ Core collapse  in  systems with equal masses,  from N-body
integrations by Makino (1996b).  The logarithm  of the central density
is plotted against time, scaled in proportion to the initial half-mass
relaxation time.  The successive curves, which correspond to different
values  of  $N$,  have been displaced  vertically    for clarity.  For
$N\le32,000$   the  first  core     collapse is clearly   followed  by
gravothermal oscillations. }
\label{}
\end{figure}

At one time, it was  not at all certain  that a cluster could  survive
beyond  the end of core  collapse, with as a singularity characterized
by infinite central  density.  Thus, many  experts doubted whether the
study of post-collapse   clusters    had   any relevance     to    the
interpretation  of observations.  H\'enon  (1961, 1965)  showed that a
cluster without such a singularity would evolve into one that did, and
he realised that, in  a real system, a   flux of energy might  well be
supplied by  the formation and evolution of  binary stars, governing a
series of collapses and expansions of the cluster core.

Numerical simulations  using  gas and  Fokker-Planck models  show that
systems with  at least a few thousand  stars  (Goodman 1987, Heggie \&
Ramamani 1989, Breeden \etal\ 1994) follow a complicated succession of
collapses  and expansions, called  gravothermal oscillations  by their
discoverers   (Sugimoto \& Bettwieser   1983,  Bettwieser  \& Sugimoto
1984).  Quite apart from their relevance in nature, these oscillations
are interesting in their own right, as an example of chaotic dynamics.
From  this point  of view they  have  been studied by  Allen \& Heggie
(1992), Breeden \& Packard (1994), and Breeden \& Cohn (1995).

In   1995 the  genuine    occurrence of  gravothermal oscillations  in
$N$-body  systems was spectacularly  demonstrated by Makino (1996a,b).
These results confirm that   the nature of post-collapse  evolution in
$N$-body  systems  is  far more   stochastic   than in the  simplified
continuum  models  on which  so much  of  our understanding   rests at
present.

\section{ Observational evidence of core collapse }

In the eighties,  CCD observations allowed a  systematic investigation
of the inner surface brightness profiles (within $\sim$ 3\arcm) of 127
galactic  globular clusters   (Djorgovski \&  King  1986, Chernoff  \&
Djorgovski  1989,  Trager  \etal\  1995).  These  authors  sorted  the
globular  clusters into two  different   classes: (i) the  King  model
clusters,    whose  surface     brightness     profiles   resemble   a
single-component King  model with a  flat isothermal core  and a steep
envelope,   and   (ii)  the  collapsed-core  clusters,   whose surface
brightness profiles  follow an almost pure power  law with an exponent
of about --1.    In the Galaxy, about   20\% of the  globular clusters
belong to the second type, exhibiting  in their inner regions apparent
departures   from  King-model   profiles.    Consequently,  they   are
considered to have collapsed cores.

The globular cluster  M15 has long been  considered as a  prototype of
the collapsed-core   star  clusters.  High-resolution imaging   of the
centre of M15 has resolved  the luminosity cusp into essentially three
bright stars.  Post-refurbishment HST star-count data confirm that the
2.2\arcs\ core radius observed  by Lauer \etal\ (1991), and questioned
by Yanny  \etal\ (1994), is  observed  neither by Guhathakurta  \etal\
(1996) with WFPC2 data nor   by Sosin \&  King  (1996) with FOC  data.
This  surface-density  profile  clearly continues   to climb  steadily
within  2\arcs.  A maximum-likelihood method  rules out a 2\arcs\ core
at  the 95\% confidence level.  It  is not possible  to distinguish at
present between a pure power-law profile  and a very small core (Sosin
\& King 1996). Consequently, among the galactic globular clusters, M15
displays one of the best cases of  clusters caught in  a state of deep
core collapse.


\section{ Tidal tails from wide-field imaging } 

\subsection{ Tidal truncation }

In addition  to the  effects  of their internal   dynamical evolution,
globular clusters suffer strong dynamical evolution from the potential
well of their host galaxy (Gnedin \& Ostriker 1997, Murali \& Weinberg
1997).    These  external  forces speed   up   the internal  dynamical
evolution  of  these stellar systems, accelerating  their destruction.
Shocks are caused by the tidal  field of the galaxy: interactions with
the  disk, the bulge  and, somehow,  with the  giant molecular clouds,
heat up the outer regions of each star cluster.  The stars in the halo
are stripped by the  tidal field.  All  globular clusters are expected
to have already lost an important fraction of their mass, deposited in
the form of individual stars in the halo  of the Galaxy (see Meylan \&
Heggie 1997 for a review).

Recent N-body simulations of globular clusters embedded in a realistic
galactic  potential  (Oh \&   Lin  1992;  Johnston \etal\  1999)  were
performed in  order  to study the  amount  of mass  loss for different
kinds   of orbits and   different kinds  of  clusters,  along with the
dynamics and the mass  segregation  in tidal tails.  Grillmair  \etal\
(1995) in an observational analysis of star counts  in the outer parts
of a few galactic  globular clusters found extra-cluster overdensities
that they associated partly with stars stripped into the Galaxy field.

\subsection{ Tidal tails from wide-field observations }

Leon, Meylan \& Combes (2000) studied the 2-D  structures of the tidal
tails associated with 20 galactic globular clusters, obtained by using
the  wavelet transform to  detect weak  structures  at large scale and
filter the  strong   background noise for  the low   galactic latitude
clusters.  They also  present N-body simulations  of globular clusters
in orbits around   the Galaxy, in  order  to study quantitatively  and
geometrically the tidal effects they encounter (Combes, Leon \& Meylan
2000).

Their  sample clusters share different  properties or locations in the
Galaxy, with various masses and structural parameters. It is of course
necessary to have very wide field imaging observations.  Consequently,
they obtained, during the years 1996 and 1997, photographic films with
the ESO Schmidt telescope.  The field of view is 5.5\deg\ \x\ 5.5\deg\
with  a scale of  67.5\arcsec/mm.  The  filters used,  viz.   BG12 and
RG630, correspond    to    $B$ and  $R$,  respectively.     All  these
photographic films were digitalized using the MAMA scanning machine of
the Observatoire de Paris, which provides  a pixel size of 10 \micron.
The astrometric  performances of the machine  are  described in Berger
\etal\ (1991).

The next step --- identification of  all point sources in these frames
--- was performed   using  SExtractor (Bertin   \&  Arnouts  1996),  a
software dedicated  to  the automatic analysis  of astronomical images
using a multi-threshold  algorithm  allowing  good object  deblending.
The detection of  the stars was done  at a 3-$\sigma$ level  above the
background.  This software, which  can deal with  huge amounts of data
(up to 60,000  $\times$ 60,000 pixels) is not  suited for very crowded
fields like the  centers of the globular  clusters, which were  simply
ignored.  A star/galaxy separation was  performed by using the  method
of star/galaxy magnitude vs. log(star/galaxy area).

For  each  field, a  ($B$  vs.    $B-V$) color-magnitude  diagram  was
constructed,  on which a  field/cluster star  selection was performed,
following the  method of Grillmair \etal\  (1995), since cluster stars
and field  stars  exhibit different colors.   In this  way present and
past    cluster members could  be  distinguished   from  the fore- and
background  field  stars by identifying in   the CMD the area occupied
primarily by cluster stars.  The envelope  of this area is empirically
chosen so as to optimize the ratio of cluster  stars to field stars in
the relatively sparsely populated outer regions of each cluster.

\subsection{ Wavelet Analysis }

With the assumption  that the data  can be viewed  as a sum of details
with  different  typical scale  lengths,  the  next  step consists  of
disentangling these details using the space-scale analysis provided by
the Wavelet Transform (WT, cf.  Slezak \etal\ 1994; Resnikoff \& Wells
1998).  Any observational signal includes also some noise, which has a
short scale  length.  Consequently the noise  is higher  for the small
scale wavelet coefficients.  Monte-Carlo simulations were performed to
estimate the noise at  each scale and apply  a 3-$\sigma$ threshold on
the   wavelet coefficients to keep  only  the reliable structures.  In
this way it is possible to subtract the short-wavelength noise without
removing details  from the signal  which has  longer wavelengths.  The
remaining overdensities of the cluster-like stars, remaining after the
application  of the wavelength transform  analysis to the star counts,
are associated with the stars evaporated  from the clusters because of
dynamical   relaxation  and/or  tidal   stripping    by  the  galactic
gravitational field.

It is  worth emphasizing that   in  this study, the  following  strong
observational biases   were taken into account:  (i)  bias due  to the
clustering of galactic field stars; (ii) bias due to the clustering of
background  galaxies; (iii) bias due  to the  fluctuations of the dust
extinction, as observed in the IRAS 100-\micron\ map.

\begin{figure}
\centering
\includegraphics[width=.7\textwidth]{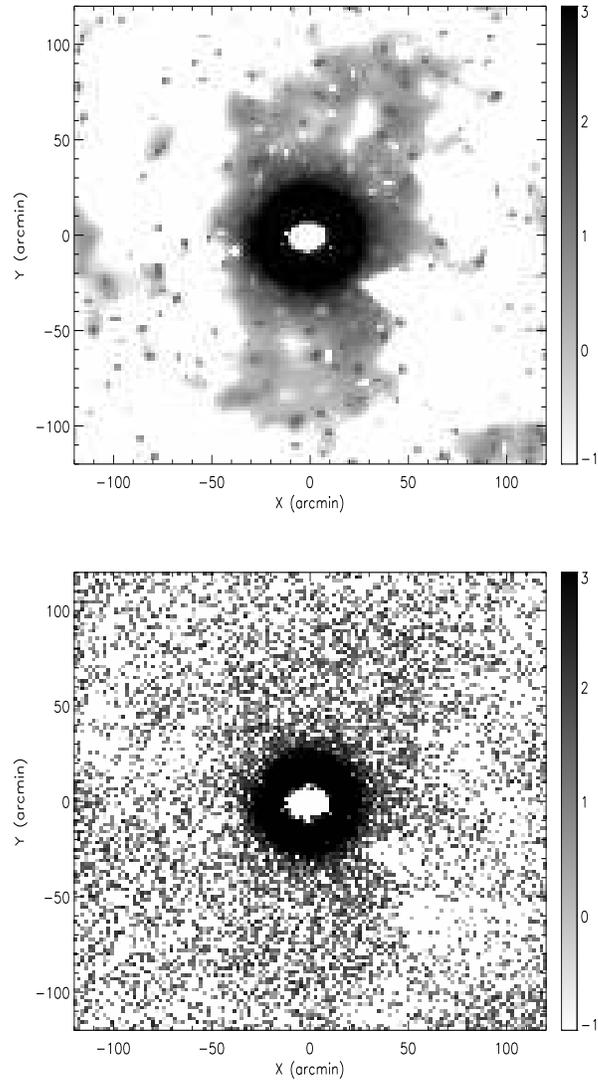}
\caption[]{NGC~5139   $\equiv$ \cent.   In   the upper panel, filtered
image  of color-selected star-count  overdensities  using the  Wavelet
Transform to be compared with the raw  star counts in the lower panel.
The upper  panel displays the full resolution  using  the whole set of
wavelet planes. From Leon \etal\ (2000). }
\label{}
\end{figure}

\subsection{ Observational Results }

The   most massive galactic globular  cluster,   \cent\ (Meylan \etal\
1995), currently crossing the disk plane, is a nearby globular cluster
located at a distance of 5.0~kpc from the sun.  Its relative proximity
allows, for the  star  count  selection,  to reach the  main  sequence
significantly below the turn-off.  Estimates,  taking into account the
possible presence  of mass segregation in its  outer parts,  show that
about 0.6 to  1~\% of its mass  has been lost  during the current disk
shocking event.  Although this cluster has, in this  study, one of the
best tail/background  S/N  ratios, it  is  by far   not the  only  one
exhibiting tidal tails.

Considering  all 20 clusters  of the sample, the following conclusions
are  reached   (see  Leon,  Meylan \&   Combes  2000  for  a  complete
description of this work):
\begin{itemize}

\item All  the clusters observed,  which  do  not  suffer from  strong
observational  biases, present  tidal  tails,  tracing their dynamical
evolution in the Galaxy (evaporation,  tidal shocking, tidal torquing,
and bulge shocking).

\item The clusters in the following sub-sample (viz. NGC~104, NGC~288,
\break NGC~2298, NGC~5139, NGC~5904,  NGC~6535, and NGC~6809)  exhibit
tidal  extensions resulting from a  recent  shock, i.e.  tails aligned
with the tidal field gradient.

\item  The clusters in  another  sub-sample (viz.  NGC~1261, NGC~1851,
NGC~1904, NGC~5694, NGC~5824,  NGC~6205, NGC~7492, Pal~5, and  Pal~12)
present   extensions which are  only tracing  the orbital  path of the
cluster with various degrees of mass loss.

\item NGC~7492 is a striking case because of  its very small extension
and  its high destruction rate   driven by the   galaxy as computed by
Gnedin \& Ostriker (1997).  Its dynamical twin  for such an evolution,
namely Pal~12, exhibits,  on the contrary,  a large  extension tracing
its orbital path, with  a  possible  shock  which happened  more  than
350~Myr ago.

\item The presence of a break in the outer  surface density profile is
a reliable indicator of some recent gravitational shocks.
\end{itemize}

Recent  CCD observations  with the Wide  Field  Imager at the  ESO/MPI
2.2-m telescope and with the CFH12K camera at the Canada-France-Hawaii
3.6-m  telescope will soon  provide improved  results, because of  the
more accurate   CCD photometry.  These  observations will   allow more
precise  observational estimates of  the mass loss rates for different
regimes of galaxy-driven cluster evolution.

\subsection{ Numerical Simulations }

Extensive numerical N-body simulations  of globular clusters in  orbit
around the Galaxy were performed  in order to study quantitatively and
geometrically the tidal effects they encounter and to try to reproduce
the above observations.   The N-body code  used  is an  FFT algorithm,
using the method of James (1977) to avoid the periodic images.  With N
= 150,000 particles, it required 2.7 seconds of CPU per time step on a
Cray-C94.

\begin{figure}
\centering
\includegraphics[width=.84\textwidth]{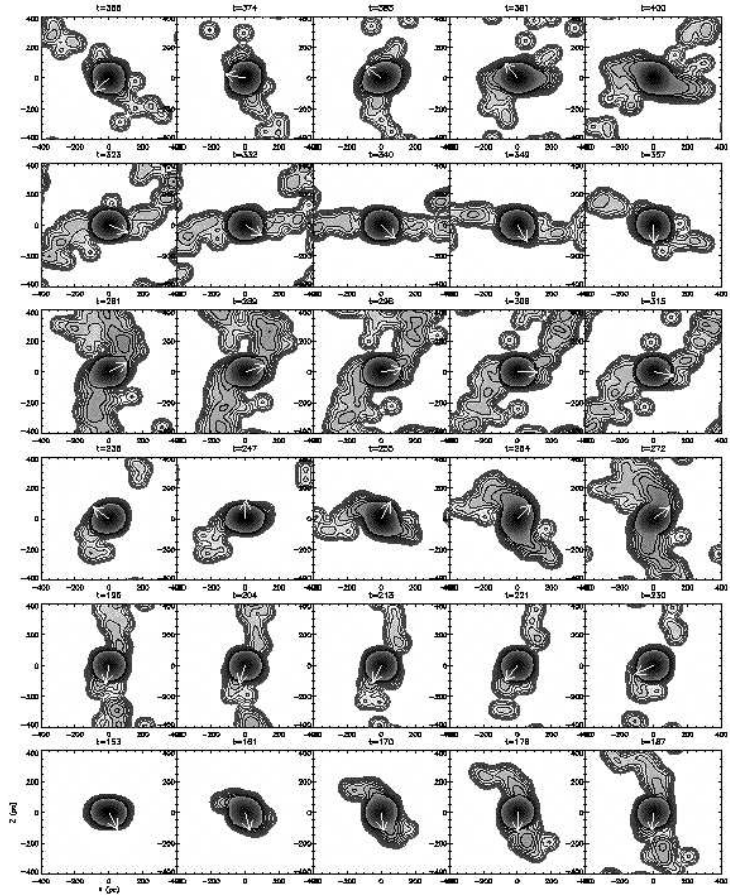}
\caption[]{  Tidal tails mapped  at different epochs  with the wavelet
algorithm  applied  to  one   of   our  simulations.    The  direction
perpendicular to the galactic  plane is indicated  by the  arrow.  The
time sequence starts with the lower-left panel and ends with the upper
right one.  The third panel exhibits tails which are quite reminiscent
of  what is observed in  NGC~5139 $\equiv$  \cent\ (see Fig.~5 above).
From Combes \etal\ (2000).}
\label{}
\end{figure}

The  globular   clusters are  represented  by  multi-mass  King-Michie
models, including mass segregation at  initial conditions.  The Galaxy
is modeled as realistically as possible, with three components, bulge,
disk and dark halo: the bulge is a  spherical Plummer law, the disk is
a Miyamoto-Nagai model, and the dark matter halo is  added to obtain a
flat galactic rotation curve.

The main conclusions of these simulations can be summarized as follows
(see Combes, Leon  \& Meylan 2000 for  a complete description  of this
work):
\begin{itemize}
\item All runs show  that the clusters  are always surrounded by tidal
tails  and debris.  This is  also true for  those that suffered only a
very slight mass loss.  These  unbound particles distribute in volumic
density like a power-law as a function of  radius, with a slope around
--4.  This  slope is much  steeper than in  the observations where the
background-foreground contamination dominates at very large scale.
\item These tails are preferentially composed of low mass stars, since
they are coming  from the external radii  of the cluster; due  to mass
segregation  built  up by   two-body relaxation,  the   external radii
preferentially gather the low mass stars.
\item For sufficiently  high and rapid mass  loss, the cluster takes a
prolate shape, whose major axis precesses around the z-axis.
\item When the tidal tail is very long (high mass loss) it follows the
cluster orbit:  the observation of the tail  geometry is thus a way to
deduce  cluster   orbits.  Stars   are  not distributed  homogeneously
through the tails, but form  clumps, and the  densest of them, located
symmetrically  in  the   tails,  are the    tracers of the   strongest
gravitational shocks.
\end{itemize}
Finally, these    N-body  experiments help   to  understand the recent
observations    of extended   tidal   tails around  globular  clusters
(Grillmair et al. 1995, Leon et al. 2000): the systematic observations
of the geometry of these tails should  provide much information on the
orbit,  dynamics, and  mass loss history  of  the clusters, and on the
galactic structure as well.

 
\section{ G1 in M31: globular cluster or dwarf galaxy~? }

The globular cluster Mayall~II $\equiv$ G1, recently observed with the
Hubble Space Telescope (HST) camera WFPC2  (Rich \etal\ 1996, Jablonka
\etal\ 1999, 2000, Meylan \etal\ 2000), is a bright star cluster which
belongs   to our  companion    galaxy, Andromeda  $\equiv$   M31.  Its
integrated visual magnitude $V$ = 13.75 mag corresponds to an absolute
visual magnitude  $M_V$  = --10.86  mag, with  $E(B-V)$  = 0.06 and  a
distance  modulus $(m-M)_{M31}$  =   24.43   mag, implying  a    total
luminosity of about $L_V$ $\sim$ 2 $\times$ $10^6 L_{\odot}$.

The coordinates of G1, viz. 
$\alpha_{G1}$(J2000.00) =  00\deg\ 32\arcm\ 46.878\arcs\ and 
$\delta_{G1}$(J2000.00) = +39\deg\ 34\arcm\ 41.65\arcs,  
when compared to the coordinates of the center of M31, viz.
$\alpha_{M31}$(J2000.00) = 00\deg\ 42\arcm\ 44.541\arcs\ and 
$\delta_{M31}$(J2000.00) = +41\deg\ 16\arcm\ 28.77\arcs, 
place it  at a projected distance of  about 3\deg, i.e. 39.5  kpc from
the center of M31.  In spite of  this rather large projected distance,
both  color-magnitude diagrams  and  radial velocities  of G1 and M31,
viz.
$V_r$(G1) = -- 331 \pmm\ 24 \kms\ while
$V_r$(M31) = -- 300 \pmm\ 4 \kms\  (21-cm HI line) and 
$V_r$(M31) = -- 295 \pmm\ 7 \kms\  (optical lines), 
completely support the idea that this  cluster belongs to the globular
cluster system of M31.

Our ($V$ vs.   $V-I$)    color-magnitude diagram reaches   stars  with
magnitudes fainter than   $V$ =  27   mag, with a   well populated red
horizontal branch at  about $V$ = 25.25 mag;  we confirm the existence
of a blueward extension of the red  horizontal branch clump as already
observed by  Rich et al.  (1996).   From model fitting, we determine a
rather high mean metallicity of  [Fe/H] = --0.95 \pmm\ 0.09,  somewhat
between  the previous  determinations    of [Fe/H] =  --0.7  (Rich  et
al. 1996) and [Fe/H] = --1.2 (Bonoli 1987; Brodie \& Huchra 1990).

From  artificial star experiments,   in  order to  estimate  our  true
measurement errors, we observe a  clear spread in our photometry  that
we attribute to an intrinsic metallicity dispersion among the stars of
G1.  Namely, adopting $E(V-I)$    = 0.10 implies a   1-$\sigma$ [Fe/H]
dispersion  of \pmm\ 0.50  dex;  adopting $E(V-I)$  =  0.05 implies  a
1-$\sigma$ [Fe/H]  dispersion of  \pmm\  0.39 dex.  In all  cases, the
intrinsic  metallicity  dispersion is    significant and  may  be  the
consequence of  self  enrichment  during the early   stellar/dynamical
evolution phases of this cluster.

\begin{figure}
\centering
\includegraphics[width=1.0\textwidth]{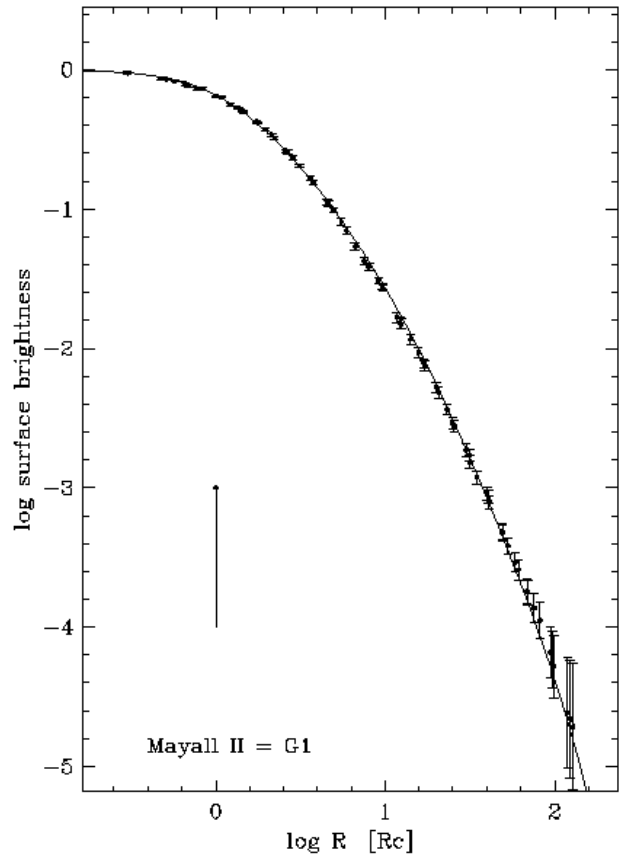}
\caption[]{  Surface    brightness profile  of  the   globular cluster
Mayall~II $\equiv$ G1, from HST/WFPC2 shallow and deep images in F555W
$\simeq$ $V$ filter;   the  continuous line represents  a  King-Michie
model (first model in Table~3)  fitted to the observed profile (Meylan
\etal\ 2000). }
\label{}
\end{figure}

We have  at our   disposal  two essential   observational  constraints
allowing the mass determination of Mayall II $\equiv$ G1:

(i)   First,  its surface  brightness   profile from HST/WFPC2 images,
providing essential structural    parameters: the core radius  \rc\  =
0.14\arcs\ = 0.52  pc, the half-mass radius  \rh\ = 3.7\arcs\ = 14 pc,
the tidal radius    \rt\  $\simeq$ 54\arcs\  =  200  pc,   implying  a
concentration \conc\ $\simeq$ 2.5 (Meylan \etal\ 2000).

(ii) Second, its central velocity  dispersion from KECK/HIRES spectra,
providing an  observed velocity dispersion \sobs\ =  25.1 \kms, and an
aperture-corrected central velocity dispersion \sigo\ = 27.8 \kms.

\subsection{ King model and Virial mass estimates }

We can  first obtain simple mass estimates  from  King models and from
the Virial  (see, e.g., Illingworth 1976).   The  first estimate, King
mass, is given by the simple equation:
$$
{\rm King~mass} ~=~ \rho_c r_c^3 \mu ~=~ 167~r_c \mu \sigma_{\circ}^2 \eqno(8)
$$
where the core radius \rc\ = 0.52 pc, the dimensionless quantity $\mu$
=  220  for  \conc\  = 2.5   (King  1966), and   the central  velocity
dispersion  \sigo\ = 27.8 \kms.   These values  determine a total mass
for the cluster of \Mtot\ = 15 $\times$  \milm\ with the corresponding
\mlv\ $\simeq$ 7.5.

The second estimate, Virial mass, is given by the simple equation:
$$
{\rm Virial~mass} ~=~ 670~r_h  \sigma_{\circ}^2 \eqno(9)
$$
where  the half-mass   radius  \rh\ =   14   pc and central   velocity
dispersion \sigo\ = 27.8 \kms. These values determine a total mass for
the  cluster  of \Mtot\ =  7.3  $\times$ \milm\ with the corresponding
\mlv\ $\simeq$ 3.6.

\subsection{ King-Michie model mass estimate }

The existing  observational constraints allow the  use of a multi-mass
King-Michie  model as defined  by Equ.~6  above.  See  \S4.1 above and
Meylan \etal\ (1995) in the case of such a model applied to \cent.  In
the case of G1, such  a model is  simultaneously fitted to the surface
brightness profile    from HST/WFPC2  and   to  the   central velocity
dispersion value from KECK/HIRES.

\begin{table}[t]
\caption{Multi-mass King-Michie models for Mayal II $\equiv$ G1}
\begin{center}
\footnotesize
\begin{tabular}{|c|c|c|c|c|c|}  
\hline
&&&&&\\
$x_{MS}^{up}$&$x_{MS}^{down}$&$M_{hr+wd}$& conc           &   \Mtot  & \mlv \\
&&&&&\\
             &             &  \%       & log ($r_t/r_c$) &  [\milm] &      \\
&&&&&\\
\hline
&&&&&\\
  1.35   &   -0.5   &  22       &      2.44       &   13.1   & 6.4   \\ 
&&&&&\\
  1.40   &   -0.2   &  21       &      2.49       &   13.9   & 6.8   \\ 
&&&&&\\
  1.40   &   +0.1   &  20       &      2.54       &   14.7   & 7.2   \\ 
&&&&&\\
  1.45   &   -0.3   &  20       &      2.48       &   14.0   & 6.9   \\ 
&&&&&\\
  1.45   &   +0.3   &  19       &      2.59       &   15.5   & 7.6   \\ 
&&&&&\\
  1.50   &   -0.4   &  20       &      2.48       &   14.1   & 7.0   \\ 
&&&&&\\
  1.50   &   +0.5   &  18       &      2.65       &   16.6   & 8.1   \\ 
&&&&&\\
  1.55   &   -0.5   &  20       &      2.47       &   14.1   & 7.0   \\ 
&&&&&\\
  1.55   &   +0.4   &  17       &      2.65       &   16.7   & 8.1   \\ 
&&&&&\\
  1.60   &   -0.3   &  19       &      2.53       &   15.0   & 7.4   \\ 
&&&&&\\
  1.60   &   +0.8   &  16       &      2.63       &   18.0   & 8.9   \\ 
&&&&&\\
\hline
\end{tabular}
\end{center}
\end{table}

An extensive grid of  about  150,000 models was  computed in  order to
explore the parameter space defined by the Initial Mass Function (IMF)
exponent $x$,  where $x$  would equal  1.35 in  the  case  of Salpeter
(1955), the   central  gravitational  potential  $W_{\circ}$,  and the
anisotropy  radius \ra.  The  IMF exponent consists  actually of three
parameters, $x_{hr}$,  describing  the heavy  remnants, resulting from
the already  evolved stars  with  initial masses in  the range between
0.85 and 100  \msun; $x_{MS}^{up}$, describing the  stars still on the
Main Sequence, with initial masses in the  range between 0.25 and 0.85
\msun; and  $x_{MS}^{down}$  describing the stars  still on   the Main
Sequence, with initial  masses in   the range  between 0.10 and   0.25
\msun.

Table~3 presents   eleven of the 50    models with the   lowest \chis,
illustrating some of the input and output parameters.  Good models are
considered as such not only on the basis  of the \chis\ of the surface
brightness  fit (see Fig.~7), but  also  from their predictions of the
observed   integrated luminosity of the    cluster  and of the   input
mass-to-light  ratio of the model.   The  different columns in Table~3
give,   for    each  model,  its    IMF  exponents   $x_{MS}^{up}$ and
$x_{MS}^{down}$; the fraction \Mhr\  of its total mass  in the form of
heavy stellar remnants  such as  neutron  stars and white  dwarfs; its
concentration \conc; its total mass   \Mtot\ of the cluster, in  solar
units; and its corresponding  mass-to-light ratio \mlv\ also in  solar
units.  Since the velocity dispersion profile is reduced to one single
value   --- the central  velocity  dispersion  ---  the models are not
strongly constrained, providing equally good  fits to rather different
sets of parameters.

The IMF exponent  $x_{hr}$,  describing the  amount of  neutron stars,
appears in all models to be very close  to $x$ = 1.35 (Salpeter 1955).
Given the lack   of   constraint from   the absence of    any velocity
dispersion profile, the  most reliable   results  are related  to  the
concentration  and the   total  mass.   With  a  concentration  \conc\
somewhere between 2.45 and 2.65, G1 presents  clearly and in all cases
the   characteristics  of a collapsed     cluster.  This is completely
different from \cent,  the  most massive  but losse  galactic globular
cluster,  which, with a  concentration of about 1.3,  has a very large
core radius  of  about 5  pc and  is  consequently very  far from core
collapse.   With a total mass  somewhere between 13  and 18 \milm, and
with the  corresponding mass-to-light ratio  \mlv\ between 6 and 9, G1
is significantly more  massive than \cent, maybe by  up to a factor of
three.  The King-Michie mass estimates are  in full agreement with the
King mass estimate, while the Virial mass estimate is smaller by about
a factor of two. It is worth mentioning that such a mass difference is
not  typical of G1:   the same factor  of  about two is also  observed
between  the  King-Michie and  Virial mass estimates   of any cluster.
See, e.g., Meylan \& Mayor (1986) and Meylan \etal\ (1995) in the case
of \cent.

\subsection{ Mayall II $\equiv$ G1 is a genuine globular cluster }

From  these    three   various mass  determinations   (King,   Virial,
King-Michie), we  can reach the following  conclusions about Mayall II
$\equiv$ G1:

(i) All mass estimates give a total mass up to three times as large as
the total mass of \cent;

(ii) With \conc\  $\simeq$ 2.5,  G1  is more concentrated  than \tuca,
which is  a massive galactic  globular cluster considered on the verge
of  collapsing;  G1 has  a  surface  brightness profile   typical of a
collapsed cluster;

(iii) G1 is the heaviest of the weighted globular clusters.

Given these  results we can wonder if,  even more than \cent, G1 could
be  a kind  of  transition step between   globular  clusters and dwarf
elliptical  galaxies.   There is  a  way of  checking this hypothesis.
Kormendy  (1985) used the  four  following quantities --- the  central
surface brightness \muo, the   central velocity dispersion  \sigo, the
core radius \rc, and  the total absolute  magnitude M --- in order  to
define  various  planes from combinations   of two  of  the above four
quantities, e.g.,  (\muo\  vs.  log~\rc).  In  all   these planes, the
various  stellar  systems plotted  by  Kormendy (1985)  segregate into
three well separated sequences: (i) ellipticals and bulges, (ii) dwarf
ellipticals, and   (iii) globular clusters.   When  plotted on  any of
these planes, G1 appears always on the  sequence of globular clusters,
and  cannot  be confused or assimilated    with either ellipticals and
bulges or dwarf ellipticals.  The same is true for \cent.

Consequently, Mayall II $\equiv$ G1 can be considered a genuine bright
and massive globular  cluster.  Actually, G1 may not  be the only such
massive globular cluster  in M31.  This  galaxy, which has about twice
as many globular  clusters  as our  Galaxy, has  at least three  other
clusters  with  central  velocity  dispersion  larger than   20  \kms\
(Djorgovski \etal\ 1997).  Unfortunately, so far,  G1 is the only such
cluster   imaged with the high   spatial  resolution of the  HST/WFPC2
camera, and consequently the    only such massive cluster   with known
structural parameters.   G1  and the  other three bright  M31 globular
clusters represent probably the high-mass and high-luminosity tails of
the otherwise  very  normal mass  and  luminosity distributions of the
rich M31 population of globular clusters.

 
\section{ Conclusion }

This review summarizes only parts of  the tremendous developments that
have   taken  place during   the   last two   decades.  These   recent
developments are  far from having  exploited all the  new capabilities
offered by the   impressive  progress in computer    simulations, made
possible by more powerful single-purpose hardware and software (Hut \&
Makino 1999, Spurzem 1998).

Observations too have still  a  lot of  information to provide,  which
will require more   elaborate modeling before  full  interpretation is
reached.  The mere observation of globular cluster stellar populations
presents some  puzzles which are  far  from being understood (Anderson
1997, 1999).

The kinematical and dynamical  understanding of globular clusters will
need the exploitation of numerous radial velocities and proper motions
of individual stars.  Only  small quantities of radial velocities have
been painfully accumulated over the last two decades, while the proper
motions have  so far simply been  ignored.   But there is  an enormous
amount of untapped information locked  in  the radial velocities  (for
one third) and proper motions (for two  thirds).  Fortunately, a large
amount of kinematical data (radial velocities and proper motions for a
few thousand stars) will soon   permit investigation of the 3-D  space
velocity distribution and     rotation in the    two largest  galactic
globular clusters, viz.  \cent\  and  \tuca\ (Freeman   \etal,  Meylan
\etal, both in preparation).


\section*{Acknowledgments}

It is a pleasure to thank my following collaborators -- 
%
T. Bridges (AAO), 
F. Combes (Paris),
G. Djorgovski (Caltech), 
P. Jablonka (Paris), 
S. Leon (Paris),
and A. Sarajedini (Wesleyan), 
%
-- for  allowing  me to  present some of   our results in   advance of
publication.


\end{document}